\documentclass[lettersize,journal]{IEEEtran}
\usepackage{amsmath,amsfonts}
\usepackage{algorithmic}
\usepackage{algorithm}
\usepackage{array}

\usepackage{textcomp}
\usepackage{stfloats}
\usepackage{url}
\usepackage{verbatim}
\usepackage{graphicx}
\usepackage{cite}
\usepackage{subfigure}
\usepackage{cleveref}
\usepackage[table,xcdraw]{xcolor}
\usepackage{balance}
\usepackage{flushend}
\usepackage{subcaption}

\begin{document}

\title{Deep Learning-based Classification of Dementia using Image Representation of Subcortical Signals}
\author{Shivani Ranjan*, Ayush Tripathi*, Harshal Shende, Robin Badal, Amit Kumar, Pramod Yadav, Deepak Joshi, and Lalan Kumar, \IEEEmembership{ Senior Member, IEEE}

\thanks{*Shivani Ranjan and Ayush Tripathi contributed equally to this work.}
\thanks{This work was supported in part by IIT Mandi iHub and HCI Foundation India with project number RP04502G.}
\thanks{This work involved human subjects or animals in its research. Approval
of all ethical and experimental procedures and protocols was granted by the
Institute Ethics Committee, All India Institute of Ayurveda, New Delhi, India, with Ref No. IEC-331/27.06.2023/Rp(E)-12/2023, dated: 17/08/2023.}
\thanks{Shivani Ranjan, Ayush Tripathi and Harshal Shende are with the Department of Electrical Engineering, Indian Institute of Technology Delhi, New Delhi - 110016, India (e-mail: eez208482@ee.iitd.ac.in; ayushtripathi1811@gmail.com, harshalsshende@gmail.com).}
\thanks{Amit Kumar and Deepak Joshi are with the Centre for Biomedical Engineering, Indian Institute of Technology Delhi, New Delhi - 110016, India (e-mail: Amit.Kumar@cbme.iitd.ac.in; joshid@iitd.ac.in).}
\thanks{Robin Badal and Pramod Yadav are with the Department of RS and BK, All India Institute of Ayurveda Delhi, New Delhi - 110076, India (e-mail: mewbadal@gmail.com; drpramod.yadav@aiia.gov.in).}
\thanks{Lalan Kumar is with the Department of Electrical Engineering,
Bharti School of Telecommunication, and,
Yardi School of Artificial Intelligence, Indian Institute of Technology Delhi, New Delhi - 110016, India (e-mail: lkumar@ee.iitd.ac.in).}

}

\markboth{Manuscript Under Review}%
{}


\maketitle

\begin{abstract}

Dementia is a neurological syndrome marked by cognitive decline. Alzheimer's disease (AD) and Frontotemporal dementia (FTD) are the common forms of dementia, each with distinct progression patterns. Early and accurate differentiation of AD and FTD is crucial for effective medical care, as both conditions present similar early symptoms, leading to frequent misdiagnosis. Traditional diagnostic methods rely on subjective screening tools, pathological biomarkers, and neuroimaging techniques such as PET and fMRI, which are time-inefficient, expensive, and less accessible. EEG, a non-invasive tool for recording brain activity, has shown potential in distinguishing AD from FTD and mild cognitive impairment (MCI). Previous studies have utilized various EEG features, such as subband power and connectivity patterns to differentiate these conditions. However, artifacts in EEG signals can obscure crucial information, necessitating advanced signal processing techniques. This study aims to develop a deep learning-based classification system for dementia by analyzing scout time-series signals from deep brain regions, specifically the hippocampus, amygdala, and thalamus. The study utilizes scout time series extracted via the standardized low-resolution brain electromagnetic tomography (sLORETA) technique. The time series is converted to image representations using continuous wavelet transform (CWT) and fed as input to deep learning models. Two high-density EEG datasets are utilized to check for the efficacy of the proposed method: the online BrainLat dataset (comprising AD, FTD, and healthy controls (HC)) and the in-house IITD-AIIA dataset (including subjects with AD, MCI, and HC). Different classification strategies and classifier combinations have been utilized for the accurate mapping of classes on both datasets. The best results were achieved by using a product of probabilities from classifiers for left and right subcortical regions in conjunction with the DenseNet model architecture. It yields accuracies of 94.17$\%$ and 77.72$\%$ on the BrainLat and IITD-AIIA datasets, respectively. This highlights the potential of this approach for early and accurate differentiation of neurodegenerative disorders.

\end{abstract}


\section{Introduction}

\subsection{Background \& Related Work}

Dementia represents a neurological syndrome impairing cognitive functioning, behaviour, and daily activities \cite{american2013diagnostic}. It leads to nerve cell degeneration and disrupted brain communication \cite{maito2023classification}.
The number of people with dementia is expected to double worldwide by 2050 \cite{scheltens2021alzheimer}, with Alzheimer’s disease (AD) being the most prevalent form, significantly contributing to this increase. Despite progress in diagnosing and managing AD, no definitive cure for AD exists. Thus, early or timely detection is a global research priority \cite{shah2016research}.

Mild cognitive impairment (MCI) is an intermediate stage between healthy ageing and dementia, with a 3–15$\%$ annual conversion rate to AD compared to 1–2$\%$ in the general population \cite{michaud2017risk}. Frontotemporal dementia (FTD), the second most common form, is characterized by changes in language, behaviour, executive function, and motor symptoms \cite{olney2017frontotemporal}. AD and FTD present similar early symptoms, often leading to misdiagnosis and complicating treatment due to their distinct progression patterns and causes \cite{musa2020alzheimer}.

Diagnostic methods face challenges due to a lack of optimal behavioural tests and the high cost of cerebrospinal fluid (CSF) and blood marker tests \cite{olsson2016csf}. Screening tools such as the Clinical Dementia Rating (CDR) \cite{morris1993clinical}, Mini-Mental State Exam (MMSE) \cite{lacy2015standardized}, Montreal Cognitive Assessment (MoCA) \cite{freitas2013montreal}, and Addenbrooke’s Cognitive Examination III (ACE-III) \cite{bruno2019addenbrooke} are useful but have limitations. These limitations include time-consuming administration, reliance on subjective judgments, influence by education level and premorbid intelligence, and less sensitivity at early stages  \cite{bruno2019addenbrooke}. There is a growing focus on identifying non-invasive brain markers to detect disease pathology before behavioural symptoms appear \cite{vrahatis2023revolutionizing}.

Mainstream early diagnosis relies on pathological biomarkers like $\beta$-Amyloid and tau Positron Emission Tomography (PET) neuroimaging \cite{jack2018nia}. AD stages are primarily associated with $\beta$-amyloid plaques and tau tangles \cite{ashrafian2021review}, while FTD involves tau or TDP-43 protein abnormalities \cite{goedert2012frontotemporal}. Imaging methods like Computed Tomography (CT), PET \cite{jack2018nia}, and functional Magnetic Resonance Imaging (fMRI) have been used in literature, with fMRI showing higher sensitivity in some cases \cite{ibrahim2021diagnostic}. Machine learning and MRI-based differentiation \cite{li20223} offer high accuracy in distinguishing these conditions \cite{sisodia2023review}. However, the practical utility of these neuroimaging methods is restricted by high infrastructure costs, less favourability in terms of patient tolerance, and brain-computer interface applications.

 Electroencephalogram (EEG) has gained significant attention as a non-invasive tool for analyzing brain activity and has proven reliable in distinguishing dementia patients from controls \cite{kongwudhikunakorn2021pilot,kim2023deep}. The suitability of EEG for repeated studies and patient monitoring makes it useful in early diagnosis and continuous tracking of AD. EEG detects changes in frequency bands, each corresponding to different functional brain alterations. These include $\delta$ (0.5-4 Hz) for slow activity, $\theta$ (4-8 Hz) for sleep-wake transitions, $\alpha$ (8-12 Hz) for resting states, $\beta$ (12-30 Hz) for attention, and $\gamma$ (above 30 Hz) for complex cognitive processes \cite{su2021constructing,cammisuli2024behavioral}. This capability aids in defining the neurophysiological profile of AD stages and differentiating it from FTD \cite{miltiadous2021alzheimer, rostamikia2024eeg}.


\begin{table*}[ht!]
\caption{A brief description of neurodegenerative disorders classification using EEG.}
\centering
\scalebox{1}{

\begin{tabular}{lllll}
\hline
\textbf{Approach}              & \textbf{Extracted Features}             & \textbf{Cases}         & \textbf{Domain} & \textbf{Reference}       \\ \hline
sLORETA and ROC                & GFP                                     & 19 FTD, 19 AD, 21 HC    & Source (Cortex)  & Nishida et al.\cite{nishida2011differences}     \\
\hline
Decision Trees, Random Forests & mean, variance, IQR, frequency          & 19 FTD, 16 AD, 19 HC   & Sensor (19)     & Miltiadous et al.\cite{miltiadous2021alzheimer} \\
\hline
DICEnet                        & band power, coherence features          & 23 FTD, 36 AD, 29 HC   & Sensor (19)     & Miltiadous et al.\cite{miltiadous2023dice} \\ \hline
Gaussian Naïve Bayes           & phase lock value, connectivity features & 23 FTD, 36 AD, 29 HC   & Sensor (19)     & Si et al.\cite{si2023differentiating}          \\ \hline
SVM                            & graph features                          & 23 FTD, 36 AD, 29 HC   & Sensor (19)     & Rostamikia et al.\cite{rostamikia2024eeg}        \\ \hline
eLORETA                        & connectivity features                   & 75 MCI, 75  HC         & Source (Cortex) & Babiloni et al. \cite{babiloni2018functional}    \\ \hline
Logistic Regression            & spectral ratios, connectivity Features  & 64 MCI, 60AD, 65 HC    & Sensor (32)     & Farina et al.\cite{farina2020comparison}     \\ \hline
Logistic Regression            & coherence, spectral power               & 53 MCI, 26 AD, 55 HC   & Sensor (20)     & Meghdadi et al.\cite{meghdadi2021resting}   \\ \hline
Extreme Learning Machine&  FuzzyEn, PLV & 28 MCI, 21 HC & Sensor (16) & Su et al.\cite{su2021constructing}\\
\hline

CEEDNet                        & EEG signals, age                        & 417 MCI, 230 AD, 459 HC & Sensor(19)      & Kim et al. \cite{kim2023deep}        \\ \hline
\end{tabular}
}
\label{T1:LS}
\end{table*}

However, artifacts from physiological and external sources can obscure or distort crucial frequency bands of EEG signals. This affects neuronal information clarity and integrity. Advancements in signal processing and the use of Machine Learning tools have improved the ability of EEG to differentiate AD from other conditions \cite{miltiadous2023dice,kim2023deep}. These improvements enhance classification accuracy and artifact removal. These tools may also aid in automation and the discovery of new neurophysiological markers \cite{komolovaite2022deep}.


Previous research on differentiating AD from FTD \cite{nishida2011differences,miltiadous2021alzheimer,miltiadous2023dice,sisodia2023review,rostamikia2024eeg} and AD from MCI \cite{babiloni2018functional,farina2020comparison,meghdadi2021resting,su2021constructing,kim2023deep} has primarily utilized EEG features such as subbands power, Global Field Power (GFP), spectral ratios, and connectivity features, as detailed in Table \ref{T1:LS}.  Despite these insights, diagnosing dementia remains challenging due to the extensive signal analysis required. Effective diagnosis requires a combination of complex features, including time-domain, frequency-domain, and connectivity metrics. As may be noted from Table \ref{T1:LS}, the current studies have primarily targeted sensor information or variations in cortical regions. However, deep brain regions, especially the hippocampus, are crucial for accurate AD and FTD classification due to their early involvement in disease progression \cite{frisoni1999hippocampal, shukla2024analyzing}. AD often begins with neurodegeneration in subcortical areas like the hippocampus before affecting the cerebral cortex \cite{smith2002imaging}. This early involvement makes deep brain regions essential for early diagnosis and precise differentiation between AD and FTD. Detecting changes in these regions can significantly enhance classification accuracy and provide earlier diagnostic insights \cite{shukla2024analyzing, quattrini2024microstructural}. Additionally, it has been established that subcortical signals can be detected using surface EEG \cite{seeber2019subcortical}. Motivated by these, the current study focuses on utilizing time-series signals from deep brain regions, specifically the hippocampus, amygdala, and thalamus for Dementia classification. 


\subsection{Objectives and Contributions}

In this work, an image representation-based framework has been presented for the classification \cite{tripathi2022imair} of three stages of dementia on two different high-density EEG datasets. In the online dataset, the 3-class classification task involves HC and subjects with FTD and AD dementia. The framework has additionally, been validated for in-house collected EEG dataset comprising of subjects with MCI, AD, and HC. The pipeline starts with the extraction of the scout time series corresponding to the left and right regions of the thalamus, hippocampus, and amygdala, using the sLORETA technique. Subsequently, the time series epochs are converted to image representation using a continuous wavelet transform. By utilizing the multi-resolution CWT-based image representation, the time-frequency maps of signals corresponding to the three categories are efficiently learned. In order to learn this mapping, the images corresponding to the left and right regions are fed to standard deep learning model architectures from the computer vision domain such as Xception \cite{xception}, ResNet \cite{resnet}, InceptionResNet \cite{inceptionresnet}, MobileNet \cite{mobilenet}, NasNetMobile \cite{nasnet}, EfficientNet \cite{efficientnet}, and DenseNet\cite{densenet}. For classifying the corresponding images, different classification strategies have been adopted. First, the models from the left and right regions are used in isolation for prediction. Subsequently, the sum and product of the posterior probabilities are utilized for the classification task. Additionally, two fusion techniques, namely, early fusion and tensor fusion networks, have also been explored for the purpose of dementia classification on both datasets. 

The remainder of the paper is organized as follows: Section II provides the description of the datasets utilized in this study (Section II.A), the scout time series extraction process  (Section II.B), preparation of image data (Section II.C), and the adopted classification strategy (Section II.D). Experimental details and results are presented in Section III, and Section IV concludes the paper.

\begin{figure*}[!t]
    \centering
    \centerline{\includegraphics[width=0.85\linewidth]{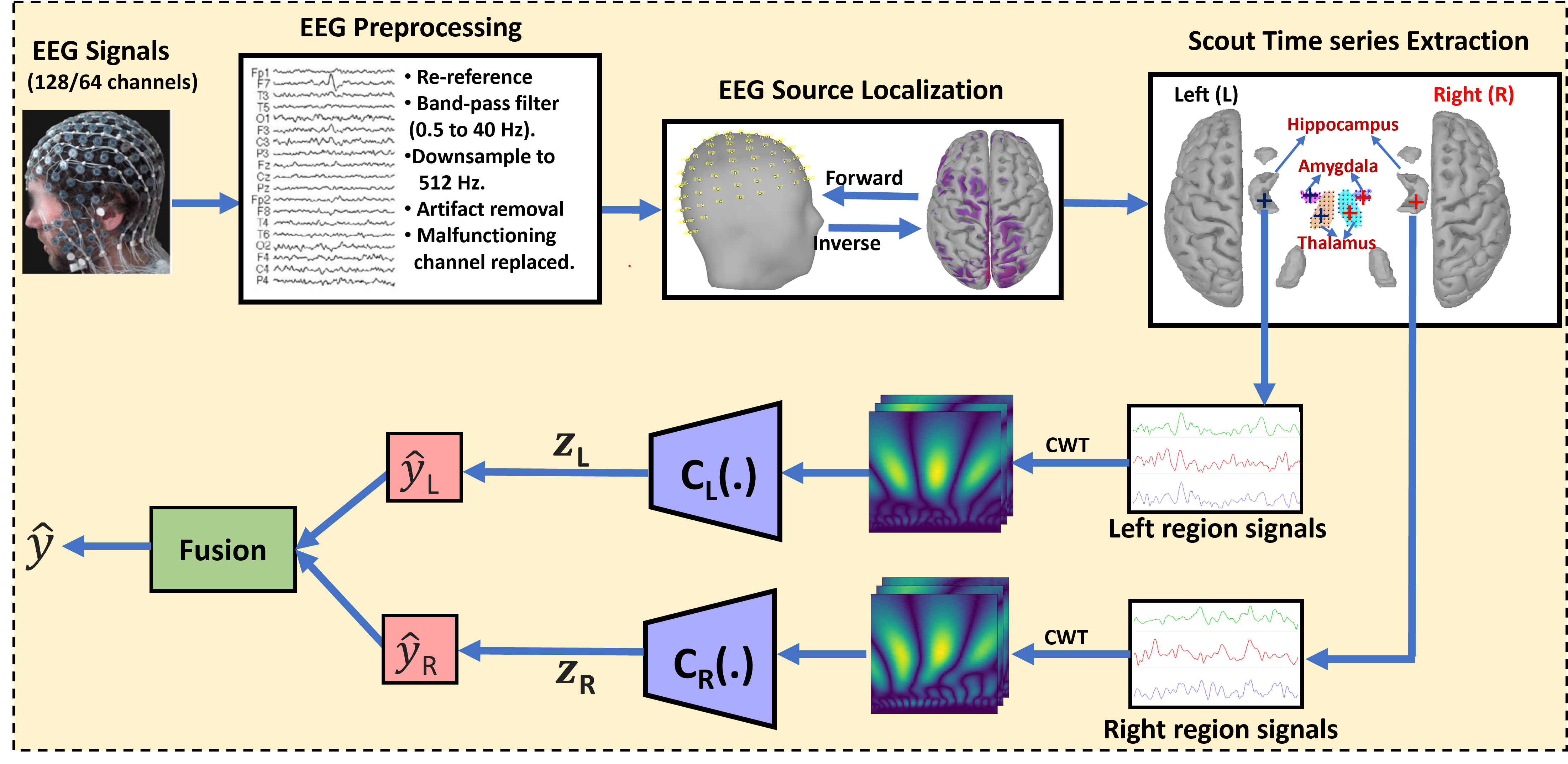}}
    \caption{Block diagram depicting the proposed method. The processed EEG signals are utilized to extract scout time series from the hippocampus, amygdala and thalamus using sLORETA. The signals are segmented and divided into left and right regions. Subsequently, the CWT-based images are fed to separate classifiers for images corresponding to left and right regions. $z_L$ and $z_R$ represent the latent representation of the classifiers, while $\hat{y}_L$ and $\hat{y}_R$ denote classifier predictions. The latent embeddings are fused using Early and Tensor Fusion, while the individual classifier outputs are fused using probability sum and product.}
    \label{fig:IM_BD}
\end{figure*}

\section{Materials and Methods}

In this section, the datasets utilized in the study, EEG preprocessing steps, scout time series extraction, image data preparation and classification strategy have been elaborated. A block diagram representing the complete pipeline is presented in Figure \ref{fig:IM_BD}.

\subsection{Dataset Description}

\subsubsection{BrainLat dataset}
The dataset utilized in this study is the section of EEG recordings released by the Latin American Brain Health Institute (BrainLat). The selected subset comprises five-minute EEG recordings from the Latin American population. More specifically, resting-state, eyes-closed recordings from 48 subjects (AD = 16; FTD = 13; HC = 19) were used for the experiments. The recordings were obtained using a 128-channel Biosemi Active II system with pin-type active, sintered Ag-AgCl electrodes referenced to contralateral linked mastoids. External electrodes were also placed periocularly to capture blinks and eye movements. Analog filters with a frequency cutoff of 0.03–100 Hz were used to reduce noise. The EEG was monitored online to detect drowsiness, muscle activity, and sweat artefacts.

The recorded data was processed offline using EEGLab \cite{delorme2004eeglab}. The first step involved in processing steps was average referencing of the EEG data. Subsequently, a bandpass filter was applied between 0.5 and 40 Hz using a zero-phase shift Butterworth filter of order 8. The data was then downsampled to 512 Hz. Independent Component Analysis (ICA) was employed to detect artefacts induced by blinking and eye movements. The components identified as artefacts were then removed to obtain clean EEG data. Malfunctioning channels were identified using a semiautomatic detection method and replaced using weighted spherical interpolation. Finally, the processed EEG signals were stored for subsequent scout time series extraction.

\subsubsection{IITD-AIIA dataset} 

The second dataset utilized in this study consists of resting-state, eyes-closed EEG data recorded from 26 (AD = 10; MCI = 8; HC = 8) right-handed participants aged 60-80 years. The data collection protocol was approved by the Institute Ethics Committee, All India Institute of Ayurveda, New Delhi. EEG data was recorded using a 64-channel Ag/AgCl active electrode EEG setup (actiCHamp, Brain Products GmbH, Germany) with Fz as the reference electrode. The signals were recorded at a sampling rate of 1000 Hz, and the 10:10 EEG electrode placement system was adopted. A conductive EEG gel was applied under each electrode to maintain resistance below 10 k$\Omega$, ensuring a high signal-to-noise ratio. No internal filters were used during the recording. The diagnosis of the AD, MCI and HC groups were based on the criteria of the MMSE \cite {lacy2015standardized} and MoCA screening tools \cite{freitas2013montreal}. Only participants whose category was consistent across both scales (MMSE: AD $<$ 18, MCI 18-25, HC $>$ 25; MoCA: AD $<$ 21, MCI 21-26, HC $>$ 26) were included in the analysis. HC group participants reported no history of neurological or psychiatric disorders. All participants provided informed consent prior to the study.

Preprocessing and analysis of EEG data were conducted using EEGLAB \cite{delorme2004eeglab} and MATLAB 2022b. Five-minute segments of continuous EEG data were bandpass filtered between 0.5 Hz and 40 Hz using a fourth-order Butterworth filter to remove irrelevant noise and enhance the signal of interest. ICA was implemented within EEGLAB to visually identify and remove components associated with blinks, eye movements, and muscle artefacts. The preprocessed data were re-referenced to an average reference. Finally, the EEG data for each subject were downsampled to 512 Hz for further scout time series extraction.

\begin{figure*}[!t]
    \centering
    \centerline{\includegraphics[width=0.85\linewidth]{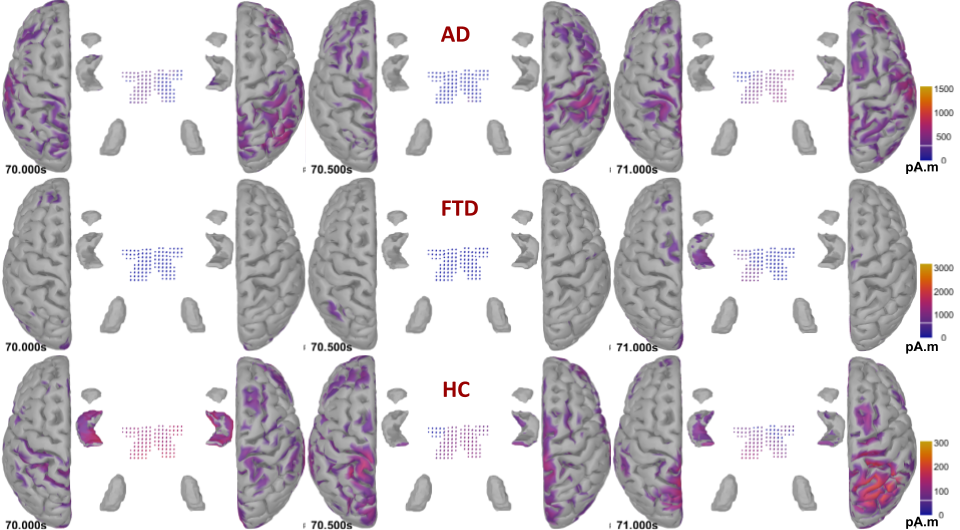}}
    \caption{Grand Average EEG Source Localization plots (front view) for AD, FTD, and HC cases from the BrainLat dataset at timestamps 70s, 70.5s, and 71s. These plots are generated using the Brainstorm Toolbox. The activation maps were set to 20$\%$ amplitude, with the amplitude threshold parameter set to "Maximum: Global" for each case. }
    \label{fig:Activation}
\end{figure*}

\subsection{Scout Time Series Extraction}

EEG source localization aims to identify the primary brain current sources that generate the measured scalp potentials \cite{michel2019eeg}. This process involves solving both the forward and inverse problems. The disparity between the number of EEG channels (128 or 64) and the number of current dipoles to be estimated (approximately 30,020) renders the source localization problem severely underdetermined. Nevertheless, in literature, it has been reported that source localization can be reasonably accurate with 64/128 channels \cite{song2015eeg}.

\subsubsection{Forward problem}
The forward problem defines the relationship between cortical currents and scalp potentials through a lead field matrix \cite{michel2019eeg}. This matrix models the propagation of currents through head tissues using Neumann and Dirichlet boundary conditions \cite{hallez2007review}. This relationship can be mathematically expressed as:

\[
V = A \tilde{S} + Z
\]

where \( V \) represents the scalp potentials, \( A \) is the lead field matrix, \( \tilde{S} \) denotes the cortical source currents, and \( Z \) is the sensor noise matrix.

The head model was computed using the Brainstorm toolbox, which involved a mixed model of cortical and deep structures \cite{attal2009modelling, attal2013assessment}. This model included 30,020 vertices, combining 15,002 from the default cortex and 15,018 from the aseg atlas. The ICBM152 MRI template and the aseg atlas were used to compute the head model for both cortical and subcortical structures, employing OpenMEEG with default conductivity parameters. To focus on specific regions, an aseg subatlas was created, including the hippocampus (surface scout), thalamus, and amygdala (volume scouts).

\subsubsection{Inverse problem}
The inverse problem estimates cortical source currents \( \tilde{S} \) using the lead field matrix \( A \). The standard low-resolution electrical tomography (sLORETA) method 


was employed for this purpose. sLORETA assumes spatial smoothness and coherence among adjacent brain regions \cite{pascual1999low}.

The source currents are estimated by solving the following optimization problem:

\[
\min_{S} F = ||V - AS||_2 + \lambda ||S||
\]

The solution is given by:

\[
\tilde{S} = A^T \left[AA^T + \lambda H \right]^{+} V = A_{sLORETA}V
\]

where \( H \) is the average reference operator, and \( A_{sLORETA} \) is the inverse kernel relating the recorded scalp potentials \( V \) to the cortical and subcortical source current estimate \( \tilde{S} \). A sample plot of the grand average brain activation for AD, FTD, and HC cases from the BrainLat dataset is depicted in Figure \ref{fig:Activation}

The brain was parcellated into left and right regions for the hippocampus, amygdala, and thalamus using the created aseg subatlas. The constrained current signals were then computed for these regions. For each of the six regions (hippocampus, amygdala, and thalamus, bilaterally) in AD, FTD, and HC cases from the BrainLat Dataset and AD, MCI and HC from the IITD-AIIA Dataset, the sources current signal belonging to a particular region are averaged out to obtain a $6$-dimensional time series matrix denoted by $\hat{S}$. This averaged time series matrix was subsequently used for image data preparation.
The EEG source localization plots for BrainLat Dataset that depict the activation difference in the brain regions specifically hippocampus, thalamus, and amygdala for AD, FTD, and HC cases is illustrated in Figure \ref{fig:Activation}

\subsection{Image Data Preparation}

Signals corresponding to the left and right regions of the thalamus, hippocampus and amygdala were extracted using the aforementioned scout time series extraction procedure. The signals are divided into $0.25$ seconds epochs which correspond to $128$ samples. This corresponds to a $128 \times 6$ dimensional matrix $\hat{S}$, where $6$ is the number of signals (corresponding to the left and right thalamus, hippocampus and amygdala). The individual time series is then converted into corresponding image representation by using the Continuous Wavelet Transform (CWT). The underlying principle behind CWT is to provide a multi-resolution representation of the time series by varying translation and scale parameters of a mother wavelet \cite{mallat1999wavelet}.  The basis functions for CWT are obtained by scaling and shifting the mother wavelet and can be mathematically expressed as: 

\begin{equation}
    \Psi_{\sigma,\tau}(t) = \frac{1}{\sqrt\sigma}\Psi\left(\frac{t-\tau}{\sigma}\right)
\end{equation}
Here, the translation is governed by parameter $\tau$ which shifts the mother wavelet in time while $\sigma$ is the scale factor. Normalization by $\frac{1}{\sqrt\sigma}$ is done to ensure that the basis function always has unit energy. Once the basis function is defined, the CWT is computed using inner product of the signal with the basis function at different translations and scaling values. For a signal $s(t)$, this is represented mathematically as,

\begin{equation}
    W_y^\Psi[\sigma,\tau] = s(t)\cdot\Psi_{\sigma,\tau}(t) = \frac{1}{\sqrt\sigma}\int_{-\infty}^{\infty} s(t) \Psi^*\left(\frac{t-\tau}{\sigma}\right)   \,dt\ 
\end{equation}

Finally, wavelet coefficients are obtained by taking all shifts and scales of the Morlet mother wavelet. The aforementioned procedure results in a $128 \times 128 \times 6$ dimensional representation $\mathcal{I}$ of the input data $X$. This is subdivided into two parts: $\mathcal{I}_L$ and $\mathcal{I}_R$ corresponding to left and right regions, respectively, which are stored subsequently. It is to be noted that each $\mathcal{I}_L$ and $\mathcal{I}_R$ have a dimension of $128 \times 128 \times 3$.

\subsection{Classification Strategy}

\begin{figure}[!t]
    \centering
    \centerline{\includegraphics[width=0.8\linewidth]{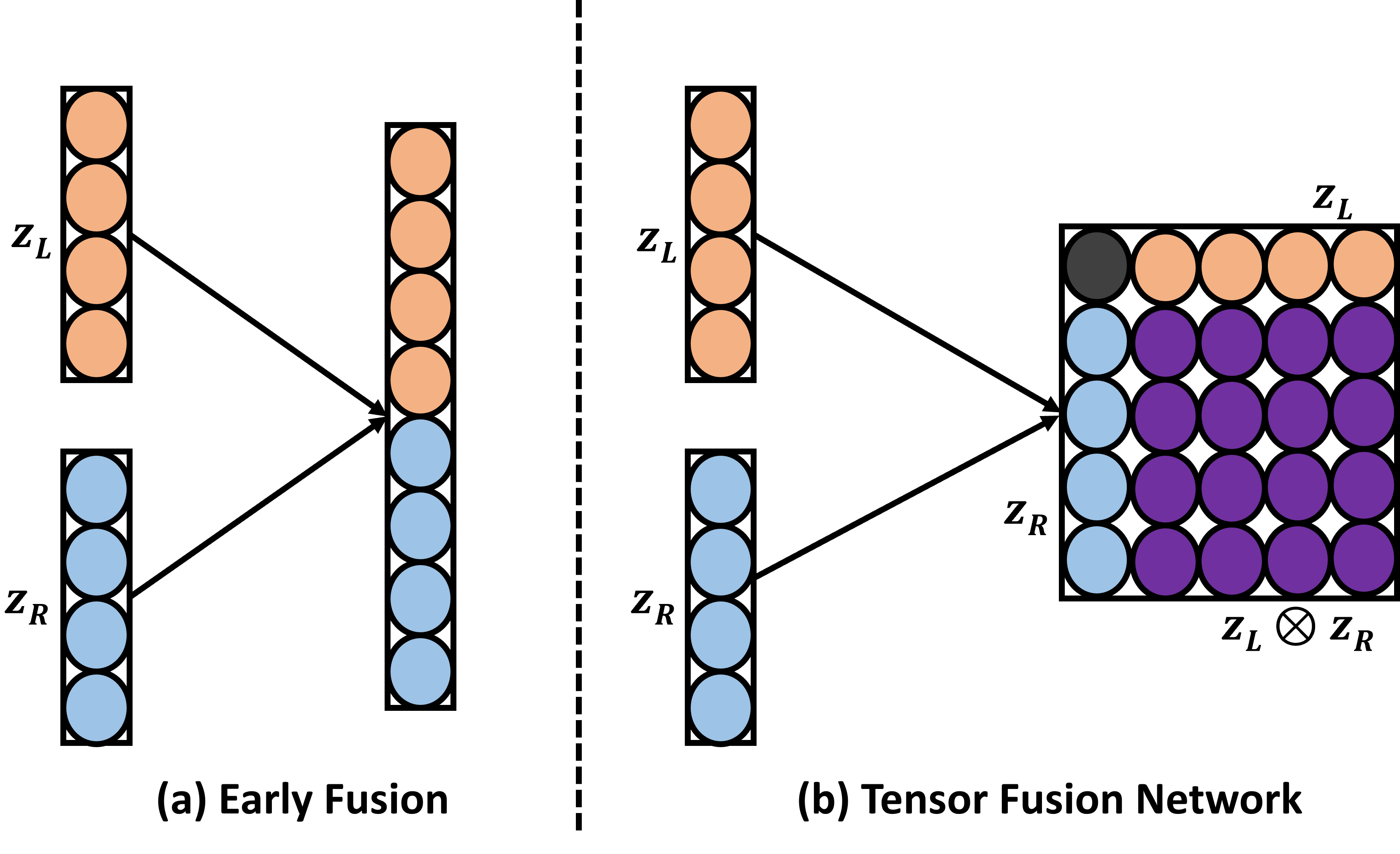}}
    \caption{Depiction of the Early Fusion and Tensor Fusion Network approaches. $z_L$ and $z_R$ represent latent embeddings from the left and right classifiers, respectively.}
    \label{fig:EF_TFN}
\end{figure}

In order to classify the images into different classes of neurodegenerative disorders (AD, FTD, HC in the BrainLat dataset and AD, MCI, HC in the IITD-AIIA dataset), several different approaches are adopted. In the first approach, two different classifier networks $C_L(.)$ and $C_R(.)$, are individually trained to map $\mathcal{I}_L$ and $\mathcal{I}_R$ to a $3$ dimensional vector representing the probabilistic distributions $P(\hat{y}_L=c_i|\mathcal{I_L};\Theta_L)$ and $P(\hat{y}_R=c_i|\mathcal{I_R};\Theta_R)$. Here, $\hat{y}_L$ and $\hat{y}_R$ is the predicted class from the left and right classifiers, and $c_i = \{AD, FTD/MCI, HC\}$ is the set of classes. $\Theta_L$ and $\Theta_R$ are the set of parameters for the classifiers corresponding to the left and right sets of images. Furthermore, a combination of posterior probabilities obtained from individual classifiers is also utilized for the classification task. The sum and product of class probabilities from individual classifiers are computed, and the input $X$ is assigned to class $\hat{y}_{sum}$ or $\hat{y}_{mul}$ based on a maximum of the computed probabilities. Mathematically, the probabilities are computed as $ P(\hat{y}_{sum}=c_i|\mathcal{I}_L,\mathcal{I}_R) = P(\hat{y}_L=c_i|\mathcal{I}_L;\Theta_L) + P(\hat{y}_R=c_i|\mathcal{I}_R;\Theta_R)$ and $P(\hat{y}_{mul}=c_i|\mathcal{I}_L,\mathcal{I}_R) = P(\hat{y}_L=c_i|\mathcal{I}_L;\Theta_L) * P(\hat{y}_R=c_i|\mathcal{I}_R;\Theta_R)$ for the sum and product cases respectively. 

Additionally, two other approaches for using the latent representations of the two classifiers are utilized for the classification task. The latent outputs of the classifiers are fused by using two different strategies: Early Fusion and Tensor Fusion Network \cite{tfn} to obtain predictions denoted by $\hat{y}_{ef}$ and $\hat{y}_{tfn}$ respectively. The two approaches have been pictorially depicted in Figure \ref{fig:EF_TFN}. It is to be noted that both early fusion and tensor fusion networks are trained in an end-to-end manner. Several different standard architectures and their variants were utilized for the classifier block, including Xception \cite{xception}, ResNet \cite{resnet}, InceptionResNet \cite{inceptionresnet}, MobileNet \cite{mobilenet}, NasNetMobile \cite{nasnet}, EfficientNet \cite{efficientnet}, and DenseNet\cite{densenet}. For all the classifiers, weights are initialized from the pre-trained models on the ImageNet task. Subsequently, Adam optimizer is used to minimize the cross-entropy loss to learn the final model parameters.

\section{Experiments and Results}

\subsection{Experimental Details}

As elaborated in Section II, the scout time series corresponding to the left and right thalamus, hippocampus and amygdala are segmented to form epochs of $0.25$ seconds. For the BrainLat dataset, this segmentation process yields a total of $15408$, $12048$, and $21648$ epochs corresponding to AD, FTD and HC, respectively. Similarly, for the IITD-AIIA dataset, a total of $11088$ AD, $10800$ MCI and $9600$ HC epochs are obtained. For both datasets, from the set of epochs, $80\%$ are randomly selected for training the deep learning models, while evaluation is done on the remaining $20\%$ epochs. Since there is a significant class imbalance in both datasets, different class weights are assigned to samples belonging to different classes based on samples in the majority class to the number of samples in a particular class. This leads to class weights of $1.405$, $1.797$, and $1$ for AD, FTD and HC classes, respectively in the BrainLat dataset. Class weights of $1$, $1.027$, and $1.155$ are assigned to AD, MCI and HC samples of the IITD-AIIA dataset. The model parameters are learnt by using the Adam optimizer while minimizing cross-entropy loss. Furthermore, at the end of each training step, the validation accuracy is monitored on a set of randomly selected $20\%$ samples from the training set. The training process is stopped if the validation accuracy does not improve over a set of $20$ continuous training steps.       

\subsection{Results and Discussion}

\begin{table*}[!t]
\caption{Recognition accuracies using different classifiers utilizing different classification strategies for the BrainLat dataset. The results for the best classification strategy for each model are depicted in blue, and the best result overall is in bold.}
\centering
\scalebox{0.95}{
\begin{tabular}{|c|c|c|c|c|c|c|}
\hline
\textbf{Model}             & \textbf{Left} & \textbf{Right} & \textbf{Early Fusion}        & \textbf{Tensor Fusion} & \textbf{Sum of Prob}         & \textbf{Product of Prob}              \\ \hline
\textbf{Xception}          & 89.19         & 90.80          & {\color[HTML]{0000FF} 91.52} & 90.69                  & 90.97                        & 90.98                                 \\ \hline
\textbf{ResNet101}         & 89.53         & 90.16          & 89.10                        & 86.68                  & 91.68                        & {\color[HTML]{0000FF} 91.68}          \\ \hline
\textbf{ResNet152}         & 88.15         & 89.62          & {\color[HTML]{0000FF} 90.94} & 90.56                  & 89.91                        & 90.08                                 \\ \hline
\textbf{InceptionResNetV2} & 87.82         & 91.95          & 90.62                        & 89.29                  & {\color[HTML]{0000FF} 92.55} & {\color[HTML]{0000FF} 92.55}          \\ \hline
\textbf{MobileNet}         & 89.10         & 91.10          & 91.02                        & 90.89                  & 92.22                        & {\color[HTML]{0000FF} 92.26}          \\ \hline
\textbf{MobileNetV2}       & 89.31         & 91.78          & 92.52                        & 90.02                  & {\color[HTML]{0000FF} 92.91} & {\color[HTML]{0000FF} 92.90}          \\ \hline
\textbf{NASNetMobile}      & 87.60         & 90.27          & 90.37                        & 90.65                  & {\color[HTML]{0000FF} 90.73} & {\color[HTML]{0000FF} 90.72}          \\ \hline
\textbf{EfficientNetB2}    & 88.73         & 91.57          & 89.97                        & 92.29                  & 92.51                        & {\color[HTML]{0000FF} 92.55}          \\ \hline
\textbf{EfficientNetB3}    & 89.61         & 90.36          & 91.33                        & 90.42                  & 92.26                        & {\color[HTML]{0000FF} 92.33}          \\ \hline
\textbf{EfficientNetB4}    & 89.46         & 91.55          & 92.15                        & 91.12                  & 92.39                        & {\color[HTML]{0000FF} 92.39}          \\ \hline
\textbf{EfficientNetB5}    & 89.27         & 90.16          & 91.44                        & 91.25                  & 91.54                        & {\color[HTML]{0000FF} 91.54}          \\ \hline
\textbf{DenseNet121}       & 90.36         & 91.25          & 92.31                        & 90.98                  & 92.68                        & {\color[HTML]{0000FF} 92.69}          \\ \hline
\textbf{DenseNet169}       & 90.70         & 91.62          & 92.63                        & 92.50                  & {\color[HTML]{0000FF} 93.53} & {\color[HTML]{0000FF} 93.52}          \\ \hline
\textbf{DenseNet201}       & 91.04         & 92.85          & 92.32                        & 91.88                  & 94.16                        & {\color[HTML]{0000FF} \textbf{94.17}} \\ \hline
\end{tabular}
}
\label{tab:Results}
\end{table*}

\begin{table*}[!t]
\caption{Recognition accuracies using different classifiers utilizing different classification strategies for the IITD-AIIA dataset. The results for the best classification strategy for each model are depicted in blue, and the best result overall is in bold. }
\centering
\scalebox{0.95}{
\begin{tabular}{|c|c|c|c|c|c|c|}
\hline
\textbf{Model}             & \textbf{Left} & \textbf{Right} & \textbf{Early Fusion}        & \textbf{Tensor Fusion} & \textbf{Sum of Prob} & \textbf{Product of Prob}              \\ \hline
\textbf{Xception}          & 69.29         & 69.02          & 73.71                        & 70.66                  & 76.06                & {\color[HTML]{0000FF} 76.61}          \\ \hline
\textbf{ResNet101}         & 68.50         & 67.39          & 72.02                        & 70.64                  & 73.50                & {\color[HTML]{0000FF} 73.85}          \\ \hline
\textbf{ResNet152}         & 66.74         & 66.35          & 71.36                        & 62.81                  & 73.15                & {\color[HTML]{0000FF} 73.52}          \\ \hline
\textbf{InceptionResNetV2} & 67.91         & 70.74          & 75.85                        & 73.88                  & 76.31                & {\color[HTML]{0000FF} 76.66}          \\ \hline
\textbf{MobileNet}         & 67.21         & 68.72          & 73.47                        & 68.42                  & 74.44                & {\color[HTML]{0000FF} 74.79}          \\ \hline
\textbf{MobileNetV2}       & 66.20         & 66.48          & 71.71                        & 67.15                  & 72.40                & {\color[HTML]{0000FF} 73.20}          \\ \hline
\textbf{NASNetMobile}      &     60.57          &       57.84         &   {\color[HTML]{0000FF} 70.02}                           &   62.49                     &     64.49                 & 65.15                                      \\ \hline
\textbf{EfficientNetB2}    & 65.88         & 64.75          & {\color[HTML]{0000FF} 74.44} & 68.24                  & 71.98                & 72.67                                 \\ \hline
\textbf{EfficientNetB3}    & 67.10         & 66.50          & 71.82                        & 69.83                  & 72.01                & {\color[HTML]{0000FF} 72.15}          \\ \hline
\textbf{EfficientNetB4}    & 67.37         & 67.89          & 71.37                        & 70.09                  & 73.02                & {\color[HTML]{0000FF} 73.53}          \\ \hline
\textbf{EfficientNetB5}    & 67.99         & 67.45          & 72.18                        & 68.23                  & 73.34                & {\color[HTML]{0000FF} 73.99}          \\ \hline
\textbf{DenseNet121}       & 69.23         & 69.55          & 74.90                        & 72.07                  & 75.10                & {\color[HTML]{0000FF} 75.52}          \\ \hline
\textbf{DenseNet169}       & 70.74         & 70.45          & 74.75                        & 70.67                  & 77.17                & {\color[HTML]{0000FF} \textbf{77.80}} \\ \hline
\textbf{DenseNet201}       & 70.55         & 71.12          & 75.36                        & 69.20                  & 77.36                & {\color[HTML]{0000FF} 77.72}          \\ \hline
\end{tabular}
}
\label{tab:Results_SHEOWS}
\end{table*}

\begin{figure*}[!ht]
\centering  
\subfigure[]{\includegraphics[width=0.4\linewidth]{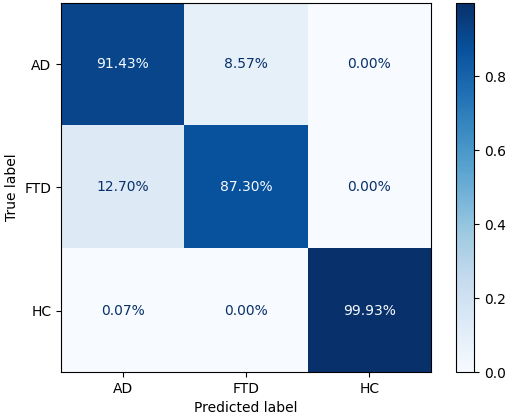}}
\subfigure[]{\includegraphics[width=0.4\linewidth]{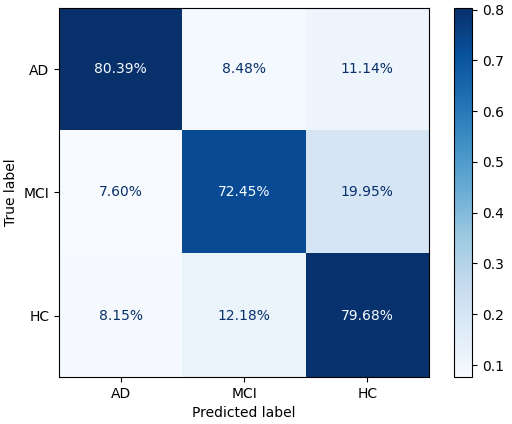}}
\caption{Confusion matrix using a combination of DenseNet201 and $\hat{y}_{mul}$ for (a) BrainLat dataset and (b) IITD-AIIA dataset.}
\label{fig:ConfMat}
\end{figure*}

\begin{figure*}[!ht]
\centering  
\subfigure[]{\includegraphics[width=0.4\linewidth]{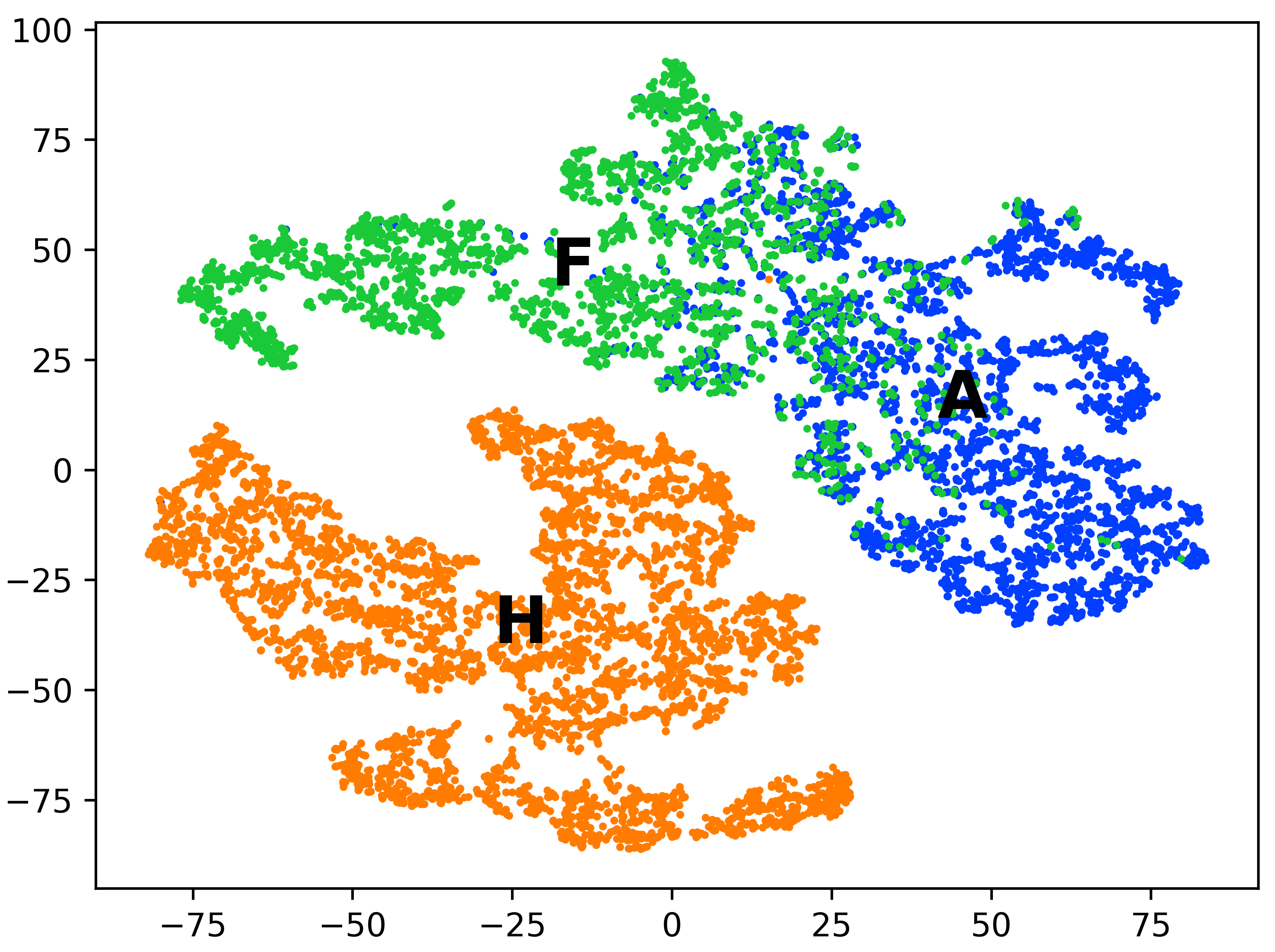}}
\subfigure[]{\includegraphics[width=0.4\linewidth]{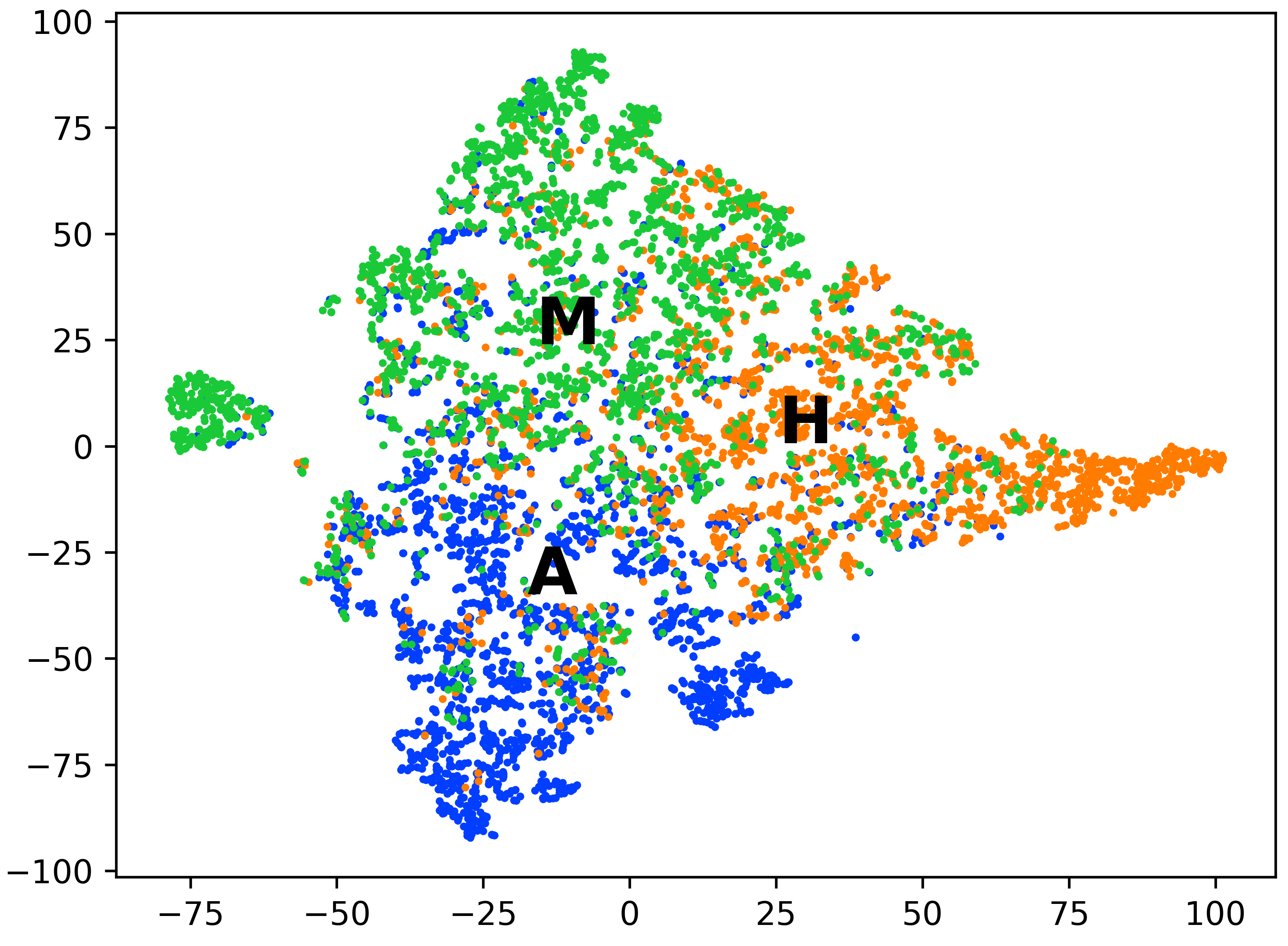}}
\caption{Scatter plot depicting clusters corresponding to each of the classes obtained by applying dimensionality reduction using t-SNE on the latent embedding vector for (a) BrainLat dataset and (b) IITD-AIIA dataset.}
\label{fig:scatterplot}
\end{figure*}

\begin{figure*}[!ht]
\centering  
\includegraphics[width=\linewidth]{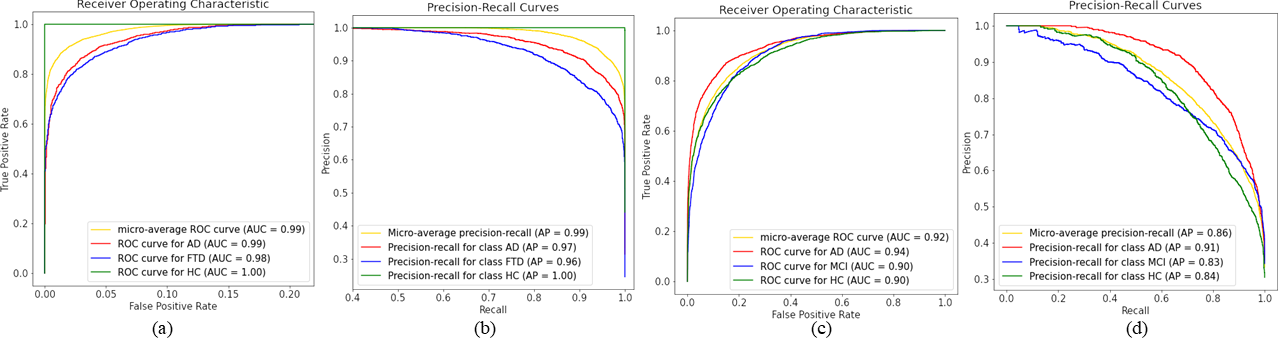}
\caption{(a), (c) Multi-class Receiver Operator Characteristic, and (b), (d) Multi-class Precision-Recall Curves for the combination of DenseNet201 and $\hat{y}_{\text{mul}}$ on the BrainLat and  datasets, respectively.}
\label{fig:ROC_PR}
\end{figure*}



The performance of the different model architectures and different approaches utilized for the task of dementia classification are presented in Tables \ref{tab:Results} and \ref{tab:Results_SHEOWS} for the BrainLat and IITD-AIIA data, respectively. The results of the individual classifiers (for left and right regions), fusion using sum and product of posterior probabilities and the latent embedding fusion approaches (using early and tensor fusion) are presented. Among the different approaches, using the product of posterior probabilities consistently yields the best classification accuracy for most of the model architectures ($12$ out of $14$ for both datasets). The DenseNet201 emerges as the best-performing model architecture, yielding accuracies of $94.17\%$ and $77.72\%$ in conjunction with the product of the probabilities approach on the two datasets. It may be observed that the classification accuracy on the IITD-AIIA dataset is low compared to the BrainLat dataset. This may be attributed to two main factors. First, the number of samples used for training the model is considerably lower in the case of the IITD-AIIA dataset. In order to learn the complex dynamics from image data, a large number of samples is required, which impacts the overall efficacy of the model. Second, the subcortical source localization in this dataset is done based on a lower number of EEG sensors ($64$ sensors). A lower number of EEG sensors leads to less accurate localization \cite{song2015eeg} of the subcortical sources, and hence, the consequent image representations lead to a comparatively lower accuracy score. Nevertheless, an accuracy of $77.72\%$ can be considered a reasonably good performance of the model architecture. Further, it is to be noted that on the BrainLat dataset, utilizing the signals from the right subcortical regions consistently leads to superior classification performance compared to the corresponding regions from the left hemisphere. This may be attributed to the fact that the right hippocampus shows more atrophy in FTD compared to the left (21$\%$ vs. 16$\%$ tissue loss). This is also consistent with previously reported findings in \cite{frisoni1999hippocampal,quattrini2024microstructural}. 

In Figure \ref{fig:ConfMat}, the confusion matrices for the three class classification tasks on both datasets are presented by using the best-performing model. It may be noted that for the BrainLat dataset, the model is particularly adept at recognizing the Healthy cases of the three classes. The majority of confusion in model predictions comes from the classification of FTD and AD classes. For the IITD-AIIA dataset, the classification accuracies of the individual classes are almost similar to each other. In order to better understand the classification heuristics, a scatter plot obtained by applying t-SNE \cite{van2008visualizing} for dimensionality reduction on the latent embedding vectors is presented in Figure \ref{fig:scatterplot} for both datasets. From the scatter plot, it may be observed that for the BrainLat dataset, the clusters corresponding to AD and FTD (depicted by A and F, respectively) have a significant overlap between them. This is particularly different from the Healthy cases (depicted by H) for which the cluster is significantly different from the other two classes. Subsequently, for the IITD-AIIA dataset, there is a significant overlap between all three clusters (AD, MCI and HC). Therefore, the misclassification trend observed in the confusion matrix is supported by the clusters corresponding to the three classes as presented in Figure \ref{fig:scatterplot}. In Figures \ref{fig:ROC_PR}, the Receiver Operator Characteristics (ROC) and Precision-Recall Curves are depicted for the BrainLat and IITD-AIIA datasets, respectively. The average curves, along with class-wise curves, are depicted in the figures. Average area under ROC values of $0.99$ and $0.92$ are obtained for the two datasets. Additionally, the average precision value for the two datasets is $0.99$ and $0.86$. The observations from the curves complement the confusion matrices and the conclusions drawn from the scatter plots.

\section{Conclusions}

In this work, a dementia classification framework using time-series signals from deep brain regions, specifically the hippocampus, amygdala, and thalamus, is presented. EEG source localization using sLORETA was leveraged to transform the average scout time series signals into image representations using CWT. The images were fed to standard model architectures from the image domain to learn the complex attributes present in the data for reliable dementia classification. The efficacy of the proposed framework was validated on two high-density EEG datasets. An online BrainLat dataset that includes subjects with AD, FTD, and HC, and an in-house collected IITD-AIIA dataset that comprises of subjects with MCI, AD, and HC, were used for the experiments. Various deep learning models, including Xception, ResNet, InceptionResNet, MobileNet, NasNetMobile, EfficientNet, and DenseNet, were used for classifying the images into one of the three categories for both the datasets. Different classification strategies, including isolated predictions from the left and right brain regions, sum and product of posterior probabilities, early fusion, and tensor fusion networks, were explored to yield optimum classification performance. The experimental results demonstrate that the proposed method achieves high classification accuracy, with the best performance observed using the combination of DenseNet201 and the product of posterior probabilities. Classification accuracy of 94.17$\%$ for the BrainLat and 77.8$\%$ for the IITD-AIIA dataset highlights the importance of focusing on deep brain regions for early and precise differentiation between AD, FTD, and MCI. This approach provides a promising baseline for future research in dementia classification and has the potential to enhance early diagnosis and treatment strategies for neurodegenerative disorders. Future work may involve increasing the sample size, exploring more sophisticated feature extraction methods from the scout time series, and employing advanced deep learning techniques to further improve system performance. Hence, the image representation-based deep learning approach has the potential to differentiate various stages of dementia, paving the way for more accurate and early diagnosis, which is crucial for the effective treatment and management of debilitating conditions.

\section{Acknowledgement}
The authors would like to thank Prof. Pradeep Kumar Prajapati, Vice Chancellor from Dr. Sarvepalli Radhakrishnan Rajasthan Ayurved University (DSSRAU), Jodhpur, and Dr. Lokesh Shekhawat, Assistant Professor from Department of Psychiatry, Atal Bihari Vajpayee Institute of Medical Sciences (ABVIMS), and Dr. Ram Manohar Lohia Hospital, New Delhi, for their expert discussion and guidance during the diagnosis and intervention processes.
Lastly, authors extend special thank to Dr. G.P. Bhagat, founder of SHEOWS Guru Vishram Vridh Ashram, for his support in facilitating the availability of the patients.

\bibliographystyle{IEEEtran}
\bibliography{mybib}

\end{document}